# Thickness Estimation of Epitaxial Graphene on SiC using Attenuation of Substrate Raman Intensity

Shriram Shivaraman[1], MVS Chandrashekhar[1], John J. Boeckl[2], Michael G. Spencer[1]
[1]School of Electrical and Computer Engineering, Cornell University, Ithaca, NY 14853
[2]Air Force Research Laboratory, WPAFB, OH 45433

**Abstract**

A simple, non-invasive method using Raman spectroscopy for the estimation of the thickness of graphene layers grown epitaxially on silicon carbide (SiC) is presented, enabling simultaneous determination of thickness, grain size and disorder using the spectra. The attenuation of the substrate Raman signal due to the graphene overlayer is found to be dependent on the graphene film thickness deduced from X-ray photoelectron spectroscopy and transmission electron microscopy of the surfaces. We explain this dependence using an absorbing overlayer model. This method can be used for mapping graphene thickness over a region and is capable of estimating thickness of multilayer graphene films beyond that possible by XPS and Auger electron spectroscopy (AES).

# Introduction

Graphene is a 2D form of carbon with a honeycomb-shaped lattice structure. It was thought to be unstable, before it was first isolated by Geim et al[1] in 2004 by mechanical exfoliation of bulk graphite. Graphene is a zero or near-zero bandgap semiconductor with a linear energy-momentum dispersion relation near the K, K' points for electrons and holes. Very unique electronic and optical properties have been predicted and/or observed for few layer graphene including extremely high carrier mobilities, room temperature quantum hall effect, band structure manipulation via electric field effect, non-vanishing minimum conductivity and plasmon amplification, to name a few[1-6]. This has led to a tremendous increase in recent years in theoretical and experimental investigation of graphene-based ultra-high speed electronic and optical devices such as field-effect transistors, pn-junction diodes, terahertz oscillators and low-noise electronic and optical sensors[6-10].

The mechanical exfoliation method of obtaining graphene requires careful selection of the cleaved pieces by optical or atomic force microscopy. Though this technique results in high quality films, it is cumbersome. Recently, epitaxial growth of graphene on SiC substrates has been demonstrated[11, 12]. This technique involves heating up the SiC substrate under vacuum to high temperatures in the range of $1200^{o}C - 1600^{o}C$. Silicon, on account of its higher vapor pressure, sublimes off more easily than carbon, leaving a carbon-rich surface behind, which rearranges on the hexagonal template provided by the substrate to form graphene. Graphene, from a few monlayers to several (>50) layers thick, can be obtained. These layers have demonstrated structural and electronic properties similar to those of graphene obtained by mechanical exfoliation including the massless Dirac-like energy dispersion relation for electrons and holes and carrier mobilities in the few tens of thousand $cm^2/V$-s range[12]. Epitaxial graphene also lends itself to optical spectroscopy studies[13, 14] which require large area layers and is believed to be a front-runner for wafer-scale technologies of graphene.

Estimating thickness of graphene grown epitaxially is an important part of its characterization. Methods for estimating thickness include Auger electron spectroscopy[11], X-ray photoelectron spectroscopy[15] and optical transmission measurements using infrared spectroscopy[16]. The first two techniques are limited by the inelastic mean free path of the auger/photo-electrons in the film (~2.1 nm for a kinetic energy of 1200eV[17]) and are limited to less than 12 monolayers. The optical transmission method is suitable only for semi-insulating substrates, as the transmission through doped substrates is very low on account of free carrier absorption.

In this paper, we present a method of estimating multilayer graphene thickness grown heteroepitaxially on SiC using Raman spectroscopy at an excitation wavelength of 488 nm. We observe a dependence between the attenuation of the substrate Raman signal and the thickness of the graphene film. We use a simple absorbing overlayer model to explain the observed dependence and extract the absorption coefficient of graphene from the fit.

The refractive index of graphene is $2.6-1.3i$[18], which corresponds to an attenuation length of ~30 nm or ~ 89 monolayers for a 488 nm wavelength optical excitation. Thus, this method can be used to estimate thickness beyond that possible by XPS and AES. Also, Raman spectroscopy is capable of estimating grain sizes and disorder in graphene[19,20]. So, it is possible to simultaneously map graphene thickness, grain sizes and disorder over a region using this technique.

## Experimental Details

4H and 6H SiC semi-insulating and $n^+$-/n-epitaxial substrates were used in the study. The diced pieces were degreased in acetone and methanol and thoroughly cleaned with a $CO_2$ snow-gun before loading them into the sublimation chamber. No special surface treatment was employed for most of these samples. Growths were conducted under high vacuum (~ $1 \times 10^{-6} - 1 \times 10^{-5}$ torr) at temperatures ranging from 1300-1600 $^oC$. Films were grown on the CMP-polished C-face of semi-insulating SiC, on the CMP-polished Si-face of $n^+$-SiC and on the Si-face of n-type (~$3 \times 10^{15} cm^{-3}$) epitaxial SiC.

## Results and Discussion

Atomic force microscopy (AFM) images of the growth on semi-insulating substrates show domain like features, which are indicative of nanocrystalline material (Fig 1(a)). n-epitaxial SiC surfaces, on the other hand, shows steplike features with wrinkles (Fig 1(b)). One particular semi-insulating sample was accidentally etched by the atmospheric oxygen in the reaction chamber and led to formation of regularly etched steps on the SiC surface. When annealed at 1400$^oC$, it gave rise to a surface morphology in Fig. 1(c), showing hexagonal faceting. This underscores the understanding that a regularly stepped surface gives rise to a high quality epitaxial graphene film. The rms roughness of the films was observed to decrease with the growth temperature.

X-ray photoelectron spectroscopy (XPS) of the grown surfaces was performed using a Surface Science Instrument SSX-100 which utilizes monochromated Aluminum K-alpha x-rays (1486.6 eV) to strike the sample surface. The X-ray beam spot was a 2mm x 1mm ellipse. Fig. 2(a) shows the high resolution C 1s peak scan for one of the samples. A number of components can be identified - SiC (282.9 eV), graphene (284.5 eV), $C_xH_y$ (285.1 eV) and C-O-H (286.2 eV)[21]. The Si 2p peak high resolution scan (Fig. 2(b)) shows two components corresponding to the substrate (100.5 eV) and $SiO_2$ (102.4 eV). Comparing the Si 2p substrate peak intensity, $I_{SiC}$, with the graphene peak intensity, $I_G$, from the C 1s scan and using the Thickogram method[22], the thickness, $t$, of the graphene overlayer can be estimated. The Thickogram is a nomograph for solving the equation:

$$\ln\left(\frac{I_G}{I_{SiC}}\right) - \left[\left(\frac{E_G}{E_{SiC}}\right)^{0.75} - \frac{1}{2}\right]\frac{t}{\lambda \cos(\theta)} - \ln 2 = \ln \sinh\left(\frac{t}{2\lambda \cos(\theta)}\right)$$

where $E_G$, $E_{SiC}$ are kinetic energies of the photoelectrons emitted at the overlayer and substrate peaks, $\theta$ is the emission angle and $\lambda$ is the inelastic mean free path of the photoelectrons. The thickness estimated from XPS has an error margin of 20% because of the uncertainty in the inelastic mean free path[15].

Alternatively, transmission electron microscopy (TEM) was used to estimate the graphene thickness. TEM samples were prepared using a Focused Ion Beam (FIB) lift-out technique. Prior to ion milling, the samples were protected with a 5keV e-beam deposited Pt cap to preserve the initial surface integrity. The samples were then prepared by FIB milling with a Ga ion beam at 30keV to a thickness of ~1 um and then finally milled using an Ar ion beam at 500eV to remove the Ga ion damage and to obtain electron transparency for high resolution imaging. Then the samples were inserted into an FEI Titan TEM operated at 80keV. Bright-field TEM images for two of the samples, T1 and T2, are shown in Figs. 3(a) and (b).

Micro-Raman spectra (Fig. 4) were recorded using a Renishaw inVia Raman microscope with a 488 nm excitation wavelength. The Raman signals from the films showed prominent characteristic graphene peaks at ~1585 cm$^{-1}$ (G) and ~2720 cm$^{-1}$ (2D). The G peak corresponds to in-plane vibrations and the 2D peak is the result of a double resonant process[19, 23]. A disorder-induced D peak at ~ 1360 cm$^{-1}$ is also seen for some of the samples. SiC also has several overtone peaks in the 1000-2000 cm$^{-1}$ regime. The peak at ~ 1516 cm$^{-1}$ is considered to be an overtone of the L point optical phonon[24]. This peak is attenuated in intensity on account of the graphene overlayer and is the basis of our thickness estimation technique detailed below.

The remaining fraction of the substrate Raman intensity after attenuation by the graphene overlayer, *S*, is estimated from the sample's Raman spectrum by removing the background corresponding to SiC. This is done by subtracting a scaled reference spectrum of pure SiC substrate. This process is depicted pictorially in Fig. 5. The same laser power was used for each sample and the corresponding reference spectrum. The fraction, *S*, can be estimated to within ±0.02 by this method.

We plot the logarithm of the remaining fraction of the substrate Raman signal, *ln(S)*, against the thickness, *t*, of the film estimated from XPS and TEM in Fig. 6. The samples for which thickness was obtained from TEM are indicated by the labels T1 and T2 on the figure. The errors involved in the data points are indicated by error bars. We find that the points fall on a straight line with slope 0.039±0.005. This dependence can be explained using a simple absorbing overlayer model (Fig. 6). Assuming graphene thickness to be *t* monolayers, an absorption coefficient *α*, the fractional Raman signal intensity from SiC post-attenuation by graphene can be written as $S = e^{-2\alpha t}$. From the fit, we extract *α* = 0.020±0.002 per monolayer for graphene, which translates to 2.0±0.2 % per monolayer. This extracted value is close 2.3% per monolayer reported for graphene on quartz substrates[25]. It is to be noted that the XPS, TEM and Raman probes have different lateral resolutions. Thus, the absorption coefficient extracted from the fit is an approximate estimate for the true absorption coefficient of graphene.

An error analysis using the error bars in *S* and *α* reveals that the percentage error in thickness estimated using this method is high for graphene upto 5 monolayers thick (Fig. 7). Beyond that, this model predicts thicknesses within an error bar of 25%. We believe that with better calibration, this error margin can be reduced.

As a proof-of-concept, a thickness mapping was undertaken for an inhomogeneous epitaxially grown graphene sample using the above method. Fig. 8 plots the variation in thickness estimated using the Raman technique over a line-scan of ~18 µm. The shaded region represents the error margins in the thickness estimates. The image at the bottom depicts the variation in G peak intensity over the same region. It is seen that a thicker graphene film corresponds to a more intense G-peak signal.

## Conclusion

In conclusion, epitaxial graphene films have been grown on SiC substrates. AFM, XPS, TEM and Raman spectroscopy have been used for characterization of the films. A simple, non-invasive and convenient method has been outlined for the estimation of the thickness of graphene layers by using the attenuation of the Raman signal from the SiC substrate. This method is capable of estimating thickness of multilayer graphene films beyond that possible by XPS and Auger electron spectroscopy (AES), as it is not limited by the small inelastic mean free path of auger/photo-electrons. Though the method, as we have presented it, has significant error for graphene films upto 5 monolayers thick, we believe that this limitation can be overcome by better calibration. This method can prove to be useful for mapping graphene thickness, grain size and disorder simultaneously over a large region.

## Acknowledgement

The authors acknowledge support from the Air Force office of Scientific research contract No. FA9550-07-1-0332 (contract monitor Dr. Donald Silversmith), and Cornell Material Science and Engineering Center (CCMR) program of the National Science Foundation (cooperative agreement 0520404).

## Figures and Figure Captions:

Fig 1: AFM micrographs of the surface taken after sublimation for some representative samples: (a) semi-insulating SiC at 1500°C with no surface preparation (z-scale 10 nm) (b) n-epitaxial SiC at 1600°C (z-scale 30 nm) (c) semi-insulating SiC at 1400°C with a regularly stepped surface prior to growth (z-scale 10 nm).

Fig. 2: XPS high resolution scans for C 1s and Si 2p peaks for one of the samples with thickness ~10 monolayers.

Fig. 3: Bright-field TEM images for two samples. (a) T1: ~2 monolayers of graphene (b) T2: ~5 monolayers of graphene.

Fig. 4: Micro-Raman spectra for (a) SiC substrate (b) ~5 monolayers epitaxial graphene on SiC (c) ~10 monolayers epitaxial graphene on SiC. The attenuation of the SiC Raman signal can be seen.

Fig. 5: Schematic depicting the subtraction procedure by which the substrate Raman signal fraction, *S*, is obtained.

Fig. 6: Substrate Raman signal fraction versus estimated thickness of graphene. Solid line corresponds to the fit.

Fig. 7: Schematic for the Raman signal attenuation model.

Fig. 8: Percentage error in the thickness estimated using the Raman technique.

Fig. 9: Raman thickness map over an 18 μm line region. The shaded area represents the error margin in the thickness estimates. The G peak intensity along the line is shown below.

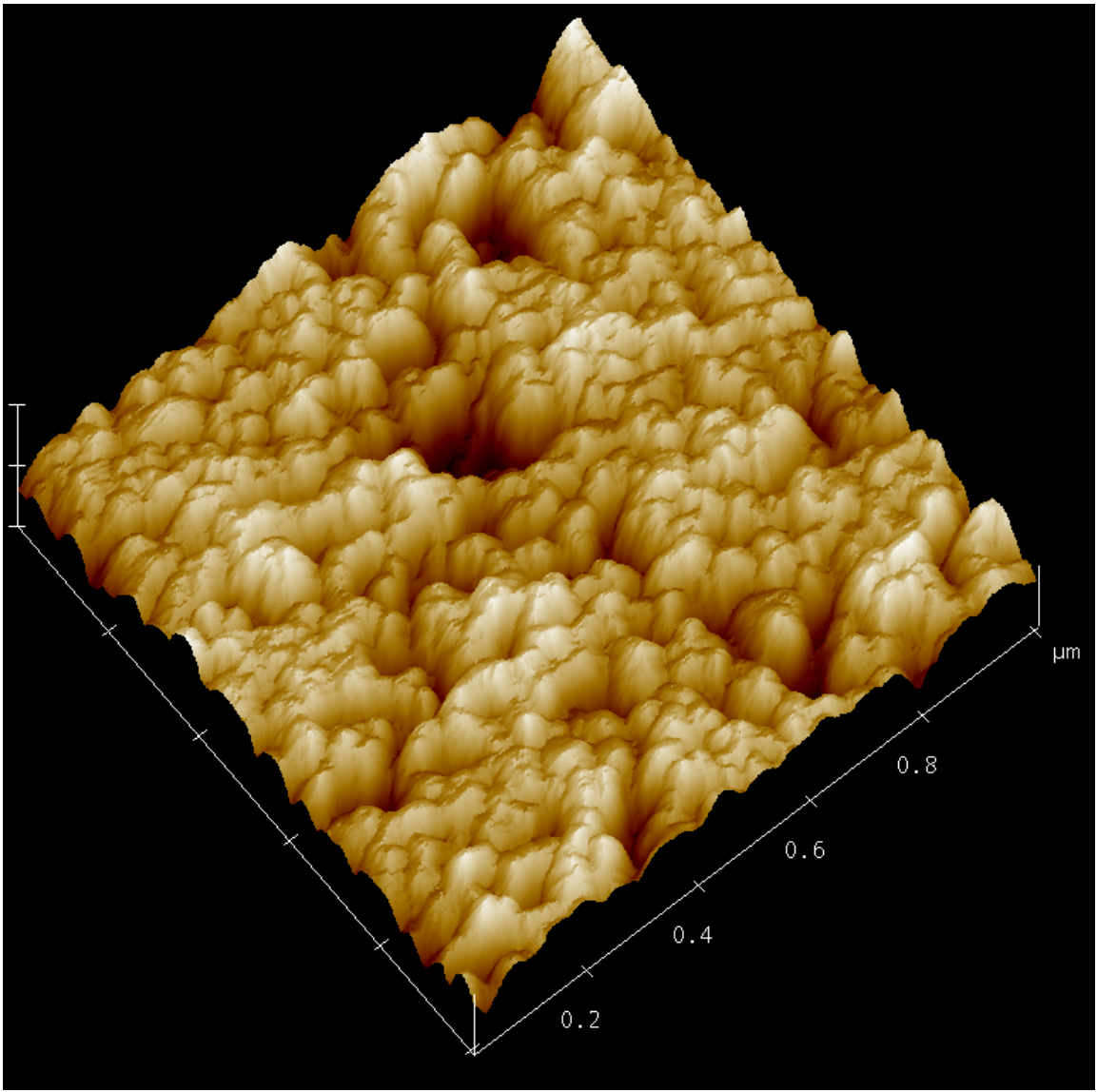

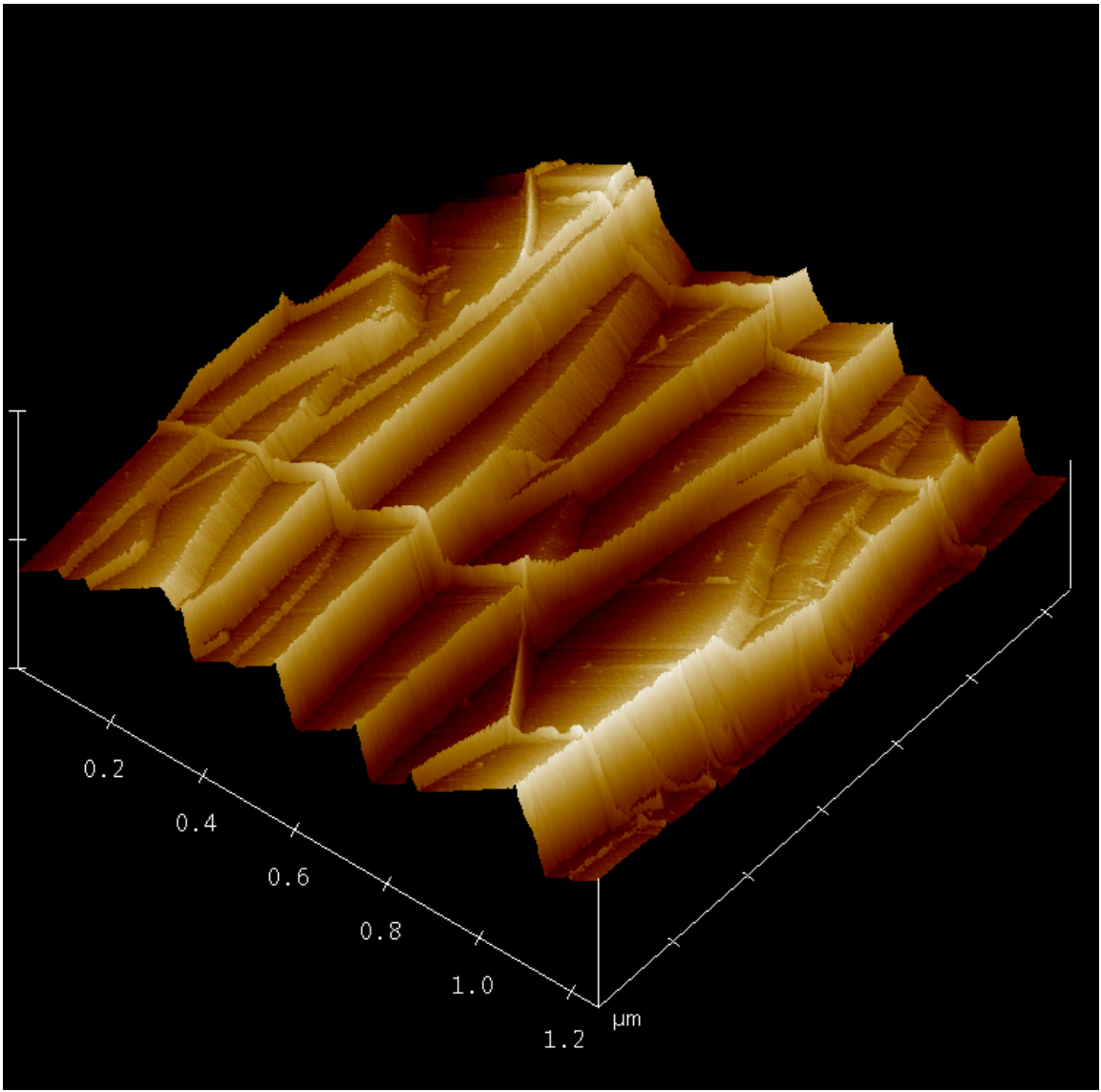

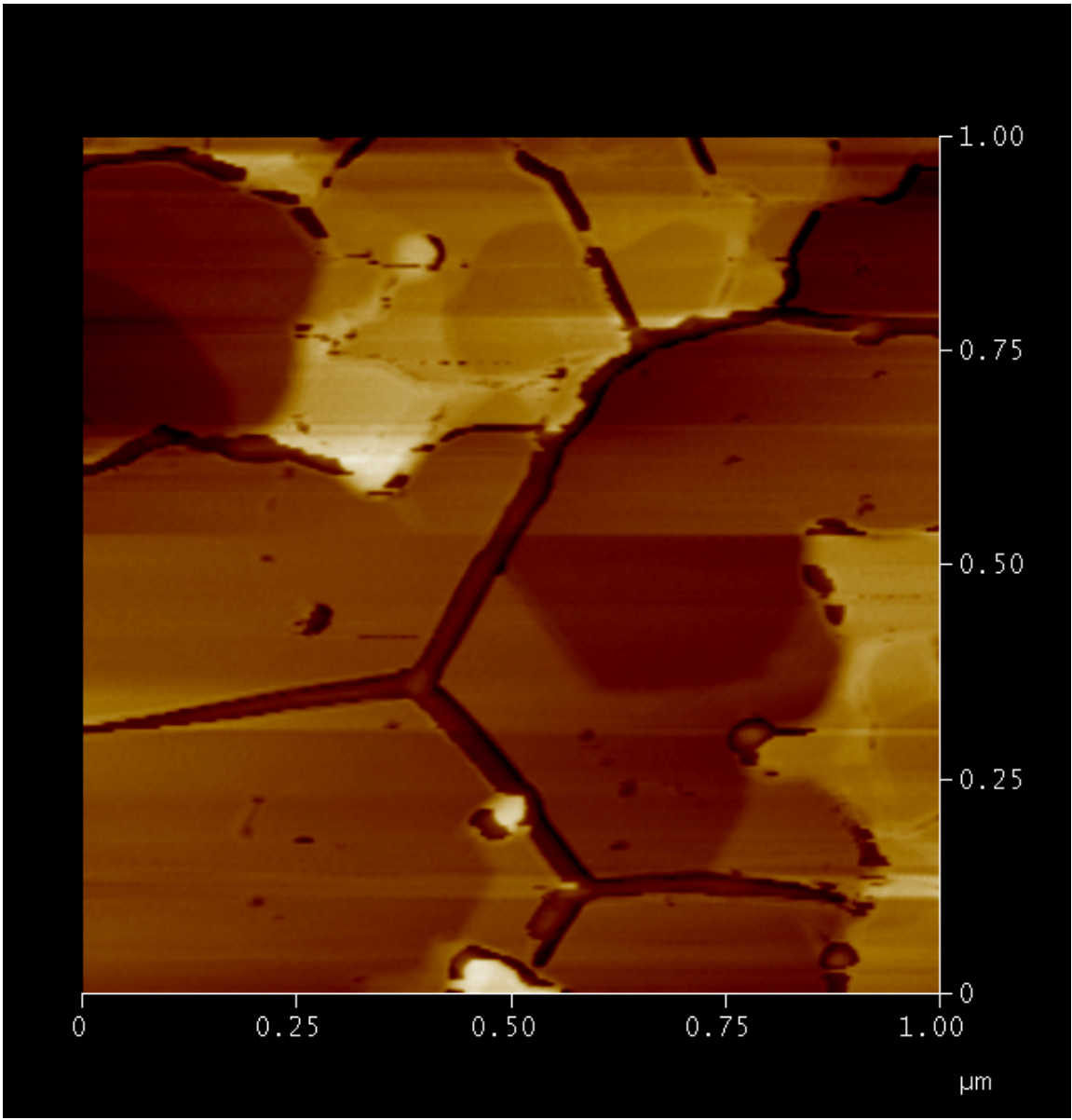

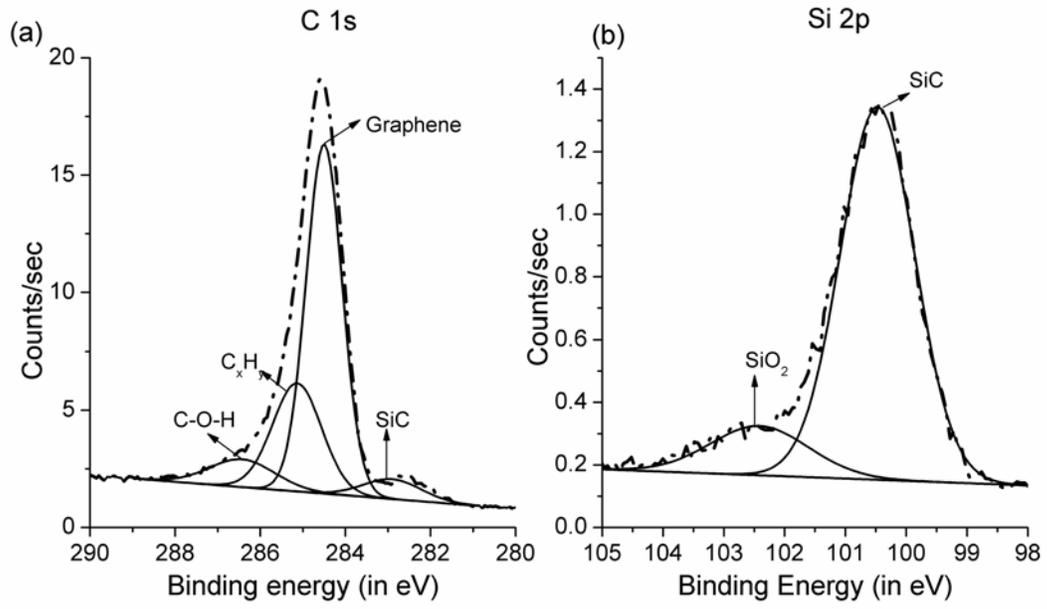

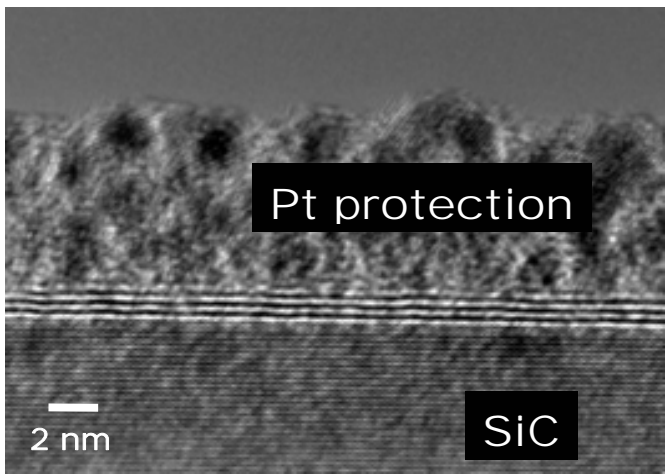
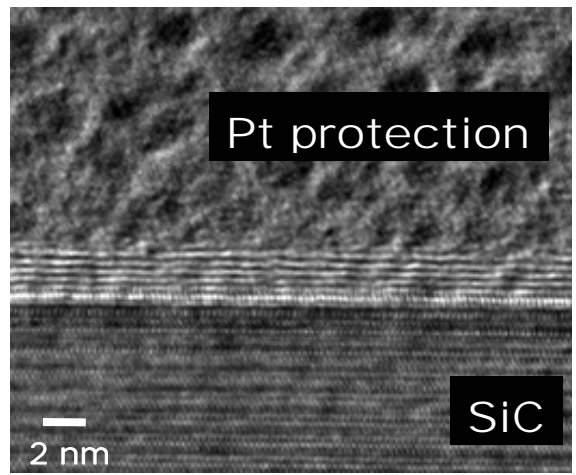

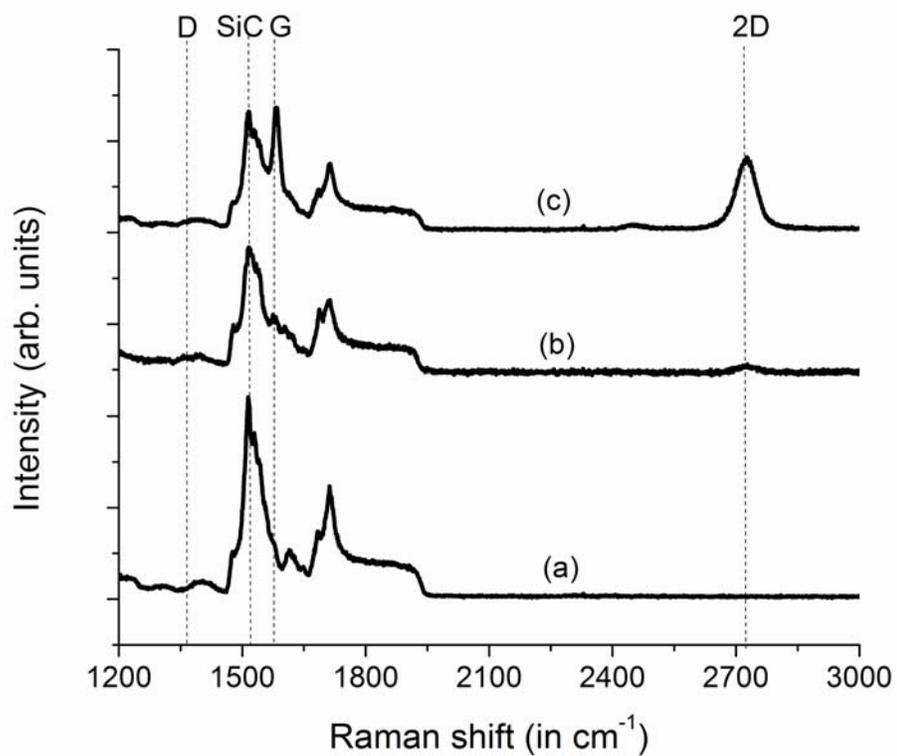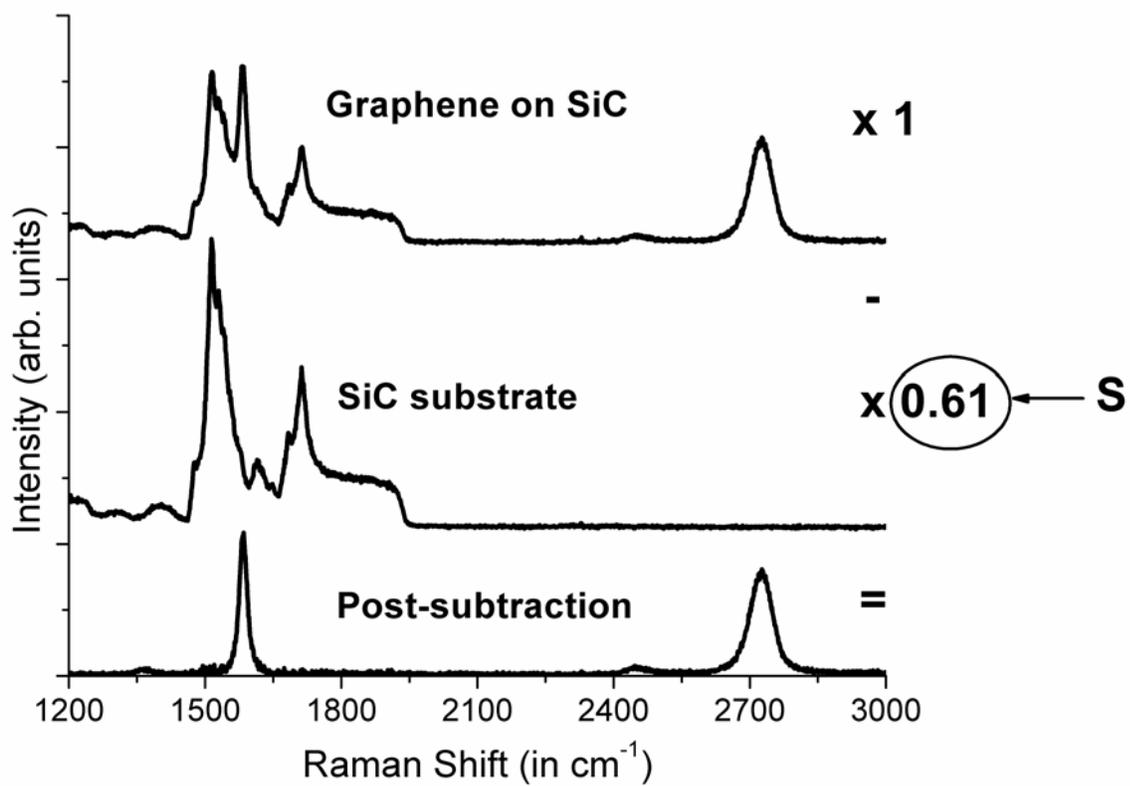

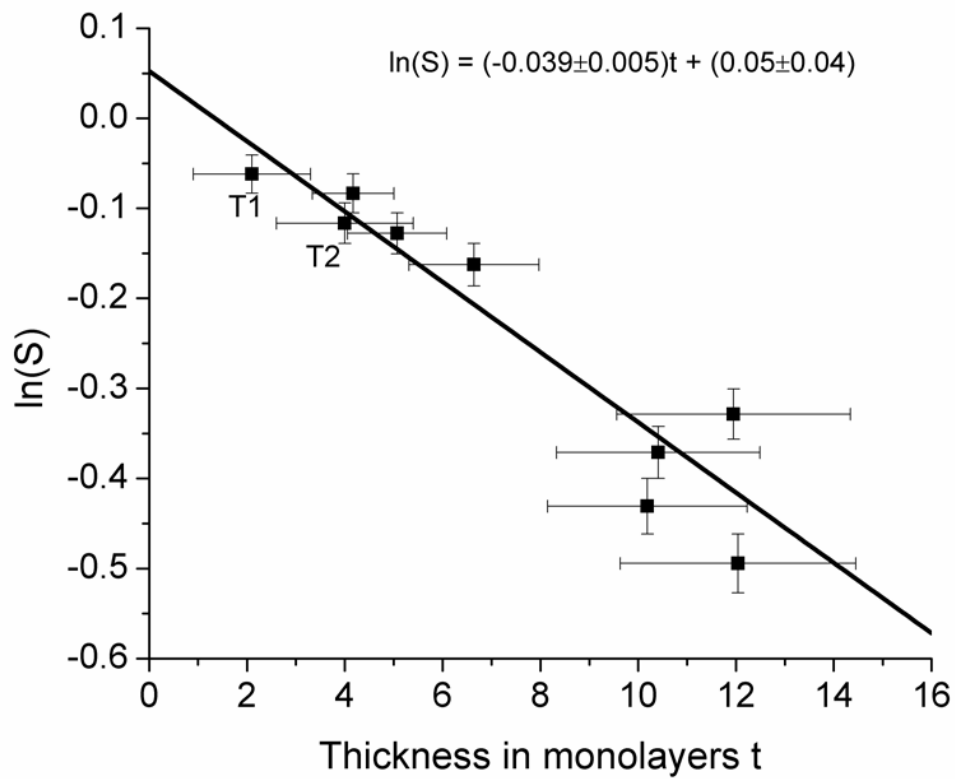

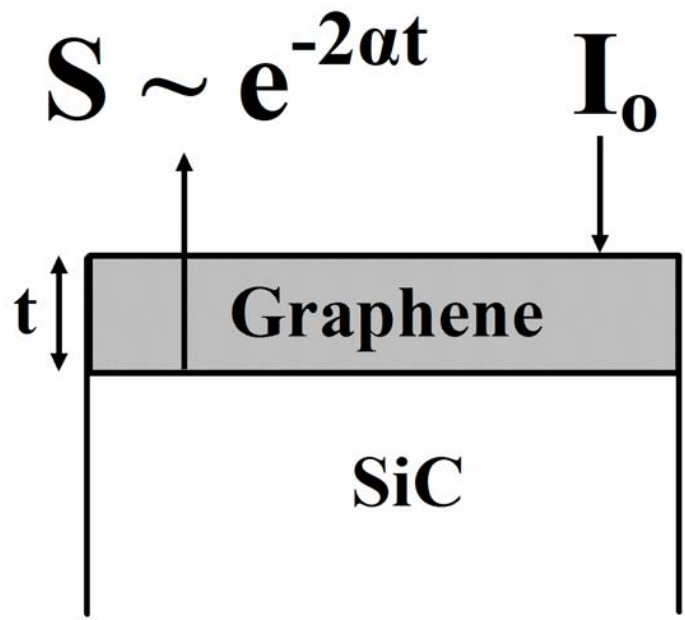

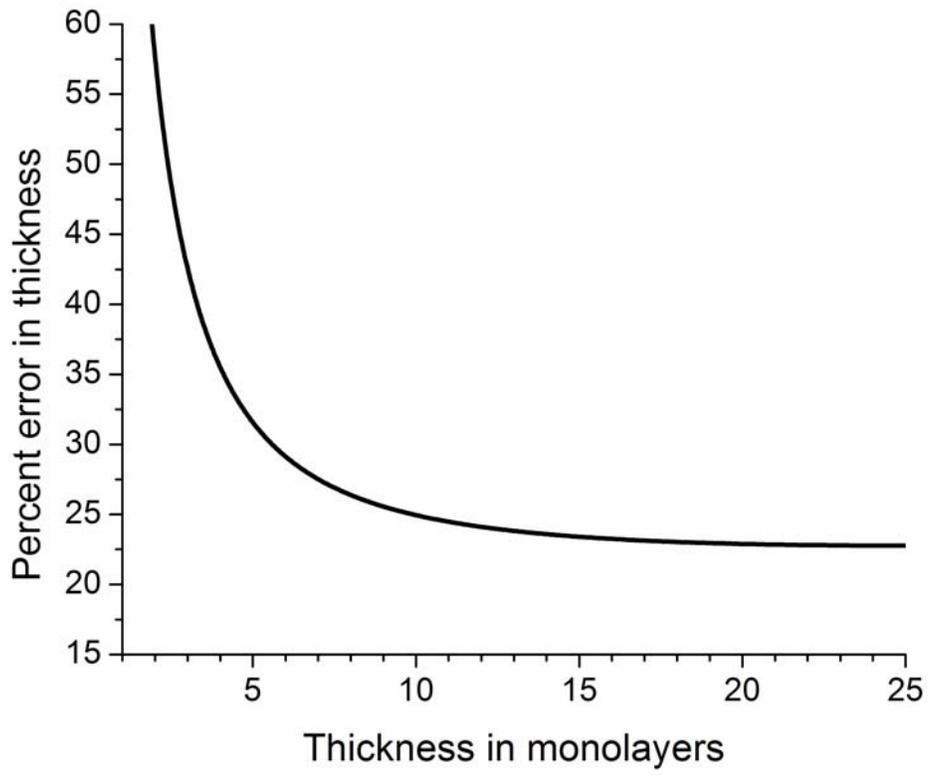
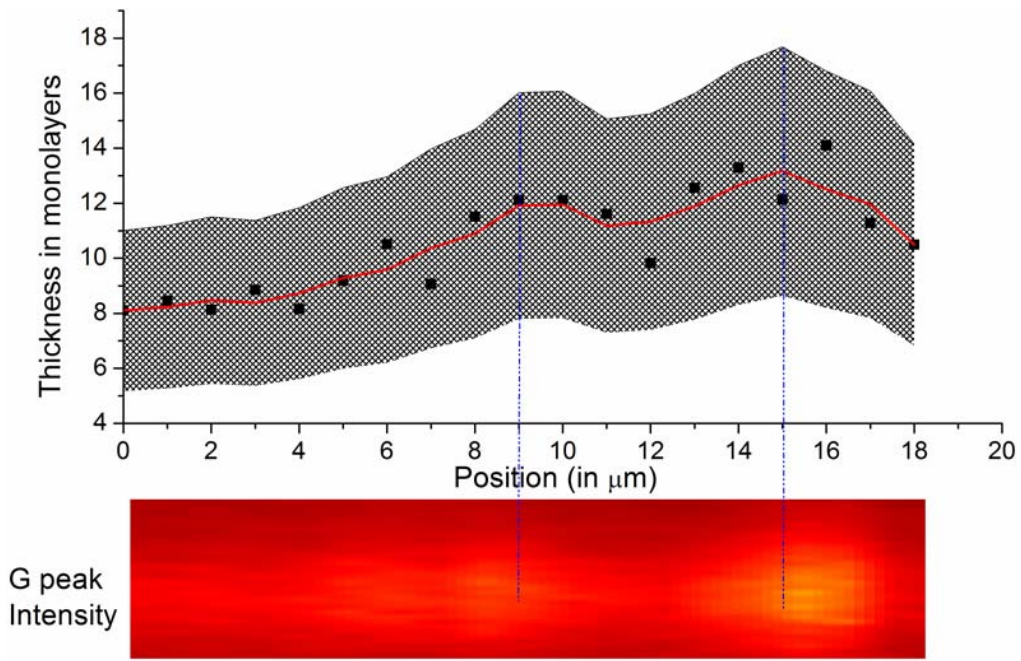